\documentclass[12pt]{iopart}

\usepackage{subcaption}
\usepackage{iopams}
\usepackage{graphicx}
\usepackage{array}
\usepackage{longtable}
\usepackage{booktabs}
\usepackage{url}
\usepackage{xurl}
\usepackage[numbers,sort&compress]{natbib}
\usepackage[hidelinks]{hyperref}

\newcommand{\asdunit}{\mathrm{m\,Hz^{-1/2}}}

\graphicspath{{figures/}}
\newcommand{\valGcrPrimaryFifty}{5.23\times10^{3}}
\newcommand{\valGcrShowerFifty}{-3.80\times10^{3}}
\newcommand{\valGcrEmissionFifty}{-113}
\newcommand{\valGcrMeanFifty}{1.31\times10^{3}}
\newcommand{\valNeutronMeanFifty}{2.38}

\newcommand{\valBackgroundMeanFifty}{1.32\times10^{3}}
\newcommand{\valSepMeanFifty}{1.12\times10^{8}}
\newcommand{\valSepMeanThousand}{3.40\times10^{6}}
\newcommand{\valBackgroundRateHundred}{1.25\times10^{3}}
\newcommand{\valStormRateHundred}{5.85\times10^{7}}
\newcommand{\valBackgroundCurrentHundred}{2.01\times10^{-16}}
\newcommand{\valStormCurrentHundred}{9.38\times10^{-12}}
\newcommand{\valCompensationReductionA}{10.7}
\newcommand{\valCompensationReductionB}{2.29\times10^{3}}
\newcommand{\valKickBackgroundOneHz}{1.71\times10^{-20}}
\newcommand{\valKickStormOneHz}{2.25\times10^{-19}}
\newcommand{\valTeBackgroundOneHz}{1.24\times10^{-20}}
\newcommand{\valTeStormOneHz}{1.06\times10^{-19}}
\newcommand{\valTeStormFiftyOneHz}{1.20\times10^{-19}}
\newcommand{\valTeStormFiftyMilliHz}{1.20\times10^{-16}}
\newcommand{\valBackgroundTotalTenthHz}{2.14\times10^{-17}}
\newcommand{\valBackgroundChargingTenthHz}{2.09\times10^{-17}}
\newcommand{\valBackgroundDischargeTenthHz}{4.00\times10^{-18}}
\newcommand{\valBackgroundKickTenthHz}{1.71\times10^{-18}}
\newcommand{\valBackgroundTotalOneHz}{3.00\times10^{-20}}
\newcommand{\valStormChargingTenthHz}{1.03\times10^{-15}}
\newcommand{\valStormDischargeTenthHz}{8.64\times10^{-16}}
\newcommand{\valStormTotalTenthHz}{1.34\times10^{-15}}
\newcommand{\valStormTotalOneHz}{1.37\times10^{-18}}
\newcommand{\valStormTotalHundredHz}{1.06\times10^{-21}}
\newcommand{\valStormTotalFiftyTenthHz}{1.89\times10^{-15}}
\newcommand{\valStormTotalThousandTenthHz}{3.14\times10^{-16}}
\newcommand{\valFzeroA}{1.33557\times10^{-4}}
\newcommand{\valLinearA}{-2.42445\times10^{3}}
\newcommand{\valQuadraticA}{6.98950\times10^{10}}
\newcommand{\valFzeroB}{1.11454\times10^{-4}}
\newcommand{\valLinearB}{-12.3607}
\newcommand{\valQuadraticB}{6.98950\times10^{10}}
\newcommand{\valForceRatio}{196.1}
\newcommand{\valRetainedForcePercent}{83.5}
\newcommand{\valSepShieldFactor}{33.0}
\newcommand{\valStormShieldFactor}{6.02}
\newcommand{\valBackgroundDriveRatio}{195.5}
\newcommand{\valStormDriveRatio}{196.1}

\newcommand{\paperrunningtitle}{Effects of radiation on the Laser Interferometer Lunar Antenna (LILA)}
\renewcommand{\leftmark}{\paperrunningtitle}
\renewcommand{\rightmark}{\paperrunningtitle}
\begin{document}
\title[\paperrunningtitle]{Effects of radiation on the Laser Interferometer Lunar Antenna (LILA)}
\author{J.~Trippe$^{1}$, B.~Shapiro$^{2}$, C.~Champagne$^{1}$, B.~Sierawski$^{1}$, M.~Hu$^{1}$, A. Sternberg, V.~Quetschke$^{3}$, T.~Creighton$^{3}$, P.~Lognonn\'e$^{4,5}$, M.~Panning$^{6}$, J.~Conklin$^{7}$, R.~Reed$^{1}$ and K.~Jani$^{1}$}
\address{$^1$ Vanderbilt University}
\address{$^2$ Johns Hopkins Applied Physics Laboratory}
\address{$^3$ University of Texas Rio Grande Valley}
\address{$^4$ Universit\'e Paris Cit\'e, Institut de physique du globe de Paris, CNRS}
\address{$^5$ California Institute of Technology}
\address{$^6$ National Aeronautics and Space Administration}
\address{$^7$ University of Florida}

\ead{james.m.trippe@vanderbilt.edu}
\begin{abstract}
The Moon provides a promising environment for gravitational-wave (GW) observations in the midband. The Laser Interferometer Lunar Antenna (LILA) combines a lunar deformation channel (LDC) and a geodesic deformation channel (GDC) to cover the midband. The LDC uses the Moon's resonant response to GW for the lower portion of the midband while the GDC measures displacement of suspended test masses to measure the latter. Radiation on the lunar surface can produce spurious signals through test-mass charging, momentum transfer, and thermoelastic deformation. We estimate these contributions for a notional 100 kg fused-silica optic using Geant4 transport, a finite-element electrostatic force map, and a spatial thermoelastic response kernel. Solar-minimum galactic cosmic rays and lunar albedo neutrons define a background case; a worst-day solar energetic particle spectrum (1989 solar flare conditions) defines an extreme, but rare, case. Aluminum-equivalent shielding from 50 to 1000 mils (1.27 to 25.4 mm) and two electrostatic actuator voltage configurations are compared. In the uniform-charge approximation, balanced drive with electrode configuration (B) substantially reduces linear charge-to-force coupling. Ideal continuous compensation of the mean charging current suppresses coherent charge accumulation while introducing negligible discharge shot noise. With compensation, charge-related noise dominates the low-frequency response, whereas thermoelastic deformation dominates at higher frequencies. Accumulated-dose and photodetector transient estimates identify additional component-qualification considerations. Estimates indicate that even under solar flare conditions, radiation noise is not a significant contributor to LILA's overall noise budget as long as charge-aware designs are followed.
\end{abstract}
\section{Introduction}
\label{sec:introduction}
The Moon offers an attractive environment for gravitational-wave (GW) detection, particularly at frequencies below the useful range of terrestrial interferometers \cite{jani2021,harms2021}. LILA is being developed around two complementary measurements: the lunar deformation channel (LDC), which measures the Moon's response to a GW, and the geodesic deformation channel (GDC), which measures relative motion of suspended test masses \cite{creighton2025,shapiro2025,quetschke2026}. The same instrumentation can address lunar interior science, including normal-mode measurements \cite{panning2025}.

The program envisages the payload-agnostic LILA Autonomous Interferometer Technology Demonstration (LILA-AITD), the 1 km initial LILA (iLILA), and the 40 km LILA Observatory. LILA-AITD is a short-duration precursor without a strain requirement. Its radiation assurance can therefore be tailored to its mission risk classification and objectives \cite{nasaclassd}. The unsuspended LDC and the suspended GDC respond differently to radiation. We only consider thermoelastic displacement as the principal direct radiation-induced noise source for the LDC. For the GDC, charge-dependent forces and particle momentum transfer can also excite the suspended optics.

The operational radiation environment contains galactic cosmic rays (GCRs), solar energetic particles (SEPs), and lunar albedo particles generated by cosmic-ray interactions in the regolith \cite{creme96,deangelis2007}. GCRs provide a persistent source of energetic protons and heavier nuclei. Large SEP events are intermittent and can greatly exceed the quiet proton intensity. We use the CREME96 worst-day prescription based on SEP fluxes averaged over the 18 h period beginning at 13:00 UT on 20 October 1989 \cite{cremesep}. Albedo neutron spectra are treated separately, with spectra taken from \cite{dobynde2021redmoon,dobynde2021redmoondata}. Although regolith interactions also produce photons and charged particles, the latter are lightly ionizing and the latter significantly range-limited \cite{deangelis2007}.

Radiation-induced disturbances have been investigated for both space-based and terrestrial GW detectors. LISA studies quantified coherent charging signals, stochastic charge noise, and particle momentum-transfer effects \cite{shaulcoherent,shaulnoise,araujo2005,lisa1998}. Cosmic-ray heating of interferometer mirrors was examined by Yamamoto and colleagues \cite{yamamoto2008}. LILA combines aspects of these problems: its suspended optics are dielectric rather than conducting LISA test masses, while its low-frequency band and exposure to the space radiation environment differ from those of terrestrial detectors.

We therefore calculate radiation transport and the detector response separately. Event-level charge, impulse and spatial energy-deposition moments are obtained from Geant4 \cite{geant42003,geant42006,geant42016}. A finite-element electrostatic map calculated with the Finite Element Method Magnetics (FEMM) \cite{femm} converts accumulated charge into longitudinal force, and a custom finite element elastic reciprocity kernel converts local heat deposition into optical-surface displacement. The main results use a spatially uniform surface-charge approximation and a notional actuator assembly inspired by LIGO \cite{advancedligo2015,ligosuspension}. Shot noise limited discharging rates are optimized against coherent charging signals without selecting a particular discharge technology. Electrode configuration is also optimized to minimize the net field around the test mass. Electronics and photodetector radiation effects are discussed separately from the mechanical and thermoelastic noise budget.

The impact of radiation on LILA's measurement sensitivity is compared to other noise sources using the methods calculated in \cite{shapiro2025,quetschke2026}. Even under the worst-day SEP conditions, these results suggest radiation noise is lower than the overall detector noise floor as long as charge-aware electrode design and discharging schemes are employed. 
\section{Simulation of the LILA GDC optic}
\label{sec:methods}
\subsection{Optic and surrounding structures}
\begin{figure}[tbp]
\centering
\includegraphics[width=\textwidth]{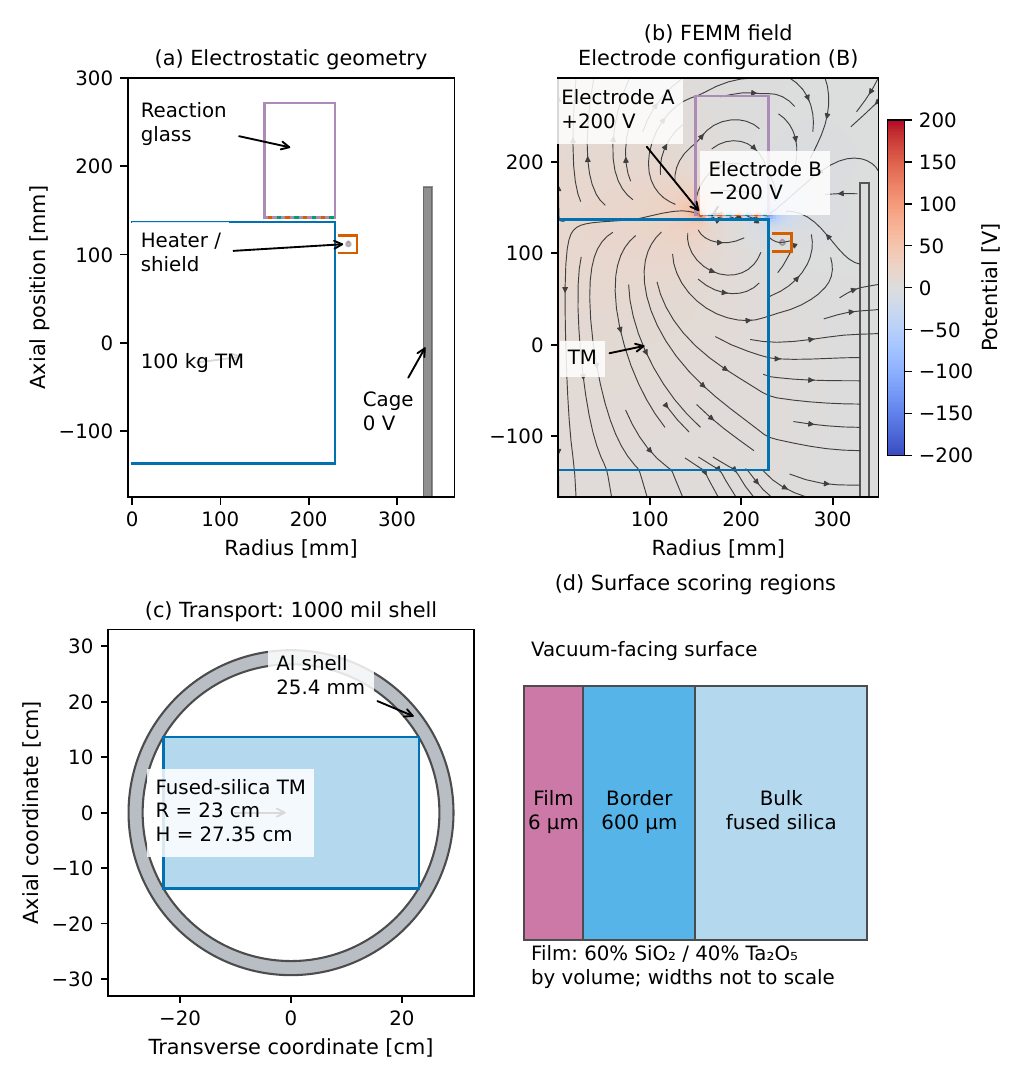}
\caption{Models used in this study. (a) The GDC electrostatic assembly, including the optic, rear electrodes, heater/shield and cage. (b) A finite-element field solution for the same assembly using electrode configuration B. (c) The simplified radiation-transport geometry: a cylindrical optic inside a spherical shell of aluminum-equivalent shielding. (d) Film, border, and bulk transport regions with different range cuts for transport fidelity, applied to both the Al shell and TM.}
\label{fig:geometry}
\end{figure}

GDC hardware is assumed in this work to be consistent with LIGO except where described here. Figure \ref{fig:geometry} shows the geometries used in simulation. The 100 kg fused-silica test mass follows the LILA optic concept of Shapiro  \cite{shapiro2025}; the surrounding electrostatic assembly is informed by Advanced LIGO actuation \cite{advancedligo2015,ligosuspension}, rather than a dimensionally exact LIGO assembly. We adopt a cylinder of radius 0.23 m, with height 0.27 m set by the total mass. The transport geometry has a uniform 6 $\mu$m outer film representing a homogenized mixture of 60\% SiO$_2$ and 40\% Ta$_2$O$_5$ by volume, with density 4.2 g cm$^{-3}$. This two-material stack is a simplified coating approximation.

The electrostatic model includes the dielectric test mass, a rear reaction-glass annulus carrying alternating electrode tracks, a heater/shield assembly and a suspension-cage surrogate. The rear gap is 5 mm; the cage is 100 mm from the optic barrel. The reaction annulus has inner and outer radii of 150 and 230 mm. Alternating concentric electrode tracks are 5 mm wide and separated by 5 mm gaps. The grounded heater/shield has a 5 mm minimum barrel clearance. The cage is represented by a finite, open-ended grounded sleeve. A more distant grounded chamber closes the numerical domain. 

We compare electrode configuration (A), with electrode potentials $+400/0$ V, and electrode configuration (B), with $+200/-200$ V. The voltage difference between adjacent electrode sets is 400 V in both cases. Electrode configuration (A) represents single-ended LIGO-like actuation with a chosen numerical reference value. Balanced drive with electrode configuration (B) reduces the common-mode field extending through the optic toward grounded surroundings, while retaining a substantial differential actuation field \cite{balancedesd}. It does not eliminate all near-electrode fields or all forces on nonuniform charge. The test mass is a dielectric, but is assumed to charge uniformly in this approximation.

Geant4 version 11.4.3 performs particle transport \cite{geant42003,geant42006,geant42016}. The reference list is FTFP\_BERT\_HP with its electromagnetic component replaced by G4EmPenelopePhysics.  This work follows the event-scoring logic used to calculate the LISA TM charging from Ara\'{u}jo et al. \cite{araujo2005}, but does not reproduce their older physics implementation exactly.

Only the optic and a pure Al spherical shell are included in the production radiation geometry. The shell's inner radius is the circumscribed radius of the cylinder; its surface therefore meets the cylinder's corners, while leaving finite clearance elsewhere. Shell thickness is varied from 50 to 1000 mil, corresponding to 1.27, 2.54, 5.08, 12.7 and 25.4 mm. This parameter represents equivalent surrounding material without needing to simulate a large world, dramatically reducing radiation transport simulation time and complexity. 

Both the TM and Al shell use the film/border/bulk layering to define material and range cuts. Production range cuts are 0.1 $\mu$m in the film, 20 $\mu$m in the border, and 600 $\mu$m by default in the bulk. The shell's inner 6 $\mu$m film region uses the same homogenized coating material as the optic - otherwise the shell is Al and the TM is fused silica within the border and bulk regions . Identical facing-surface emission properties are an idealized means of limiting mean emission imbalance, motivated by the matched-surface discussion in \cite{araujo2005}. 

\subsection{Discharging assumptions}
\label{sec:discharge}
Two discharge operations are considered. The uncompensated coherent diagnostic uses an ideal periodic reset that returns the net free charge to zero every $\tau=6$ h. For a dielectric this specifies net charge, not a unique surface potential. The nominal compensated spectra instead use a continuous signed removal rate $R_d$ that cancels the mean incident charging rate, without abrupt reset transients. Positive $R_d$ denotes delivery of electrons to cancel positive charging. Charge management has been demonstrated for conducting LISA Pathfinder test masses using UV illumination \cite{lisadischarge}; low-energy charged beams have also been studied for insulating and isolated objects \cite{chargedbeams}. Here we assume that discharging is possible and shot noise limited. 

For independent single-electron transfers at constant mean rate, the one-sided discharge-current spectrum and charge spectrum are
\begin{equation}
S_{I,d}=2e^2|R_d|,\qquad S_{Q,d}(f)=\frac{2e^2|R_d|}{(2\pi f)^2}.
\label{eq:dischargeshot}
\end{equation}
Here $e$ is the elementary charge and $f$ is Fourier frequency. Equation~(\ref{eq:dischargeshot}) contains fundamental shot noise only. Other sources of technical noise are implementation-driven and outside the scope of this study.

The choice for net-compensation is motivated by optimizing the following integral 
\begin{equation}
\mathcal J(R_d)=\int_{0.1\,\mathrm{Hz}}^{100\,\mathrm{Hz}}
\left[S_{x,\mathrm{stoch}}(f;R_d)+H_x^2(f;R_d)\right]\,\mathrm df .
\label{eq:optimization}
\end{equation}
$S_{x,\mathrm{stoch}}$ contains mechanical stochastic noise, including discharge shot noise, and $H_x$ is the coherent peak envelope defined below. For all conditions in this paper, total net discharge was the optimal choice. Further discussion is thus limited to comparing uncompensated (6 hour reset period) versus compensated (constant discharging). 

\subsection{Charge-to-force transfer from finite elements}
\label{sec:femm}
FEMM 4.2 electrostatics \cite{femm} treats the optic as a linear dielectric with relative permittivity 3.8. Free surface charge is prescribed uniformly over both end faces and the barrel. Under the assumption that Maxwell stress applies, the longitudinal force is represented by
\begin{equation}
\begin{array}{r@{\;}c@{\;}l}
F_z(Q)&=&F_0+A Q+B Q^2,\\[8pt]
A&=&\displaystyle\frac{F_z(Q_p)-F_z(-Q_p)}{2Q_p},\\[10pt]
B&=&\displaystyle\frac{F_z(Q_p)+F_z(-Q_p)-2F_z(0)}{2Q_p^2}.
\end{array}
\label{eq:forcefit}
\end{equation}
$Q_p=1$ nC was chosen for convenient normalization. The constant $F_0$ is the ordinary uncharged dielectric actuation force and is not added to radiation noise.

The coefficients for electrode configuration (A) are $F_0=\valFzeroA$ N, $A=\valLinearA$ N C$^{-1}$ and $B=\valQuadraticA$ N C$^{-2}$. For electrode configuration (B) they are $\valFzeroB$ N, $\valLinearB$ N C$^{-1}$ and $\valQuadraticB$ N C$^{-2}$. Balanced drive with electrode configuration (B) reduces $|A|$ by a factor of \valForceRatio{} while retaining \valRetainedForcePercent\% of $F_0$. The quadratic coefficient B does not vary significantly. The charge-noise coupling used here is $k_Q=A+2B\bar Q$. This is the dielectric counterpart of the charge-dependent electrostatic force used in the Shaul framework \cite{shaulcoherent,shaulnoise}.

\subsection{Source sampling and event observables}
\label{sec:transport}
Solar-minimum GCRs use the archived CREME96 quiet input with year parameter 1977 \cite{creme96,cremequiet}. The quiet model includes its prescribed low-energy and anomalous components as well as GCRs. Lunar albedo neutrons use the numerical surface distribution from REDMoon for the 2019 solar minimum, with the BON2014 GCR model, modulation potential 424 MV and the Apollo 17 regolith profile \cite{dobynde2021redmoon,dobynde2021redmoondata}. These are combined into a representative background environment. The worst-day scenario adds the CREME96 SEP result to that background. This treatment ignores other albedo particles, or enhancement of the neutron spectra from an SEP.

The REDMoon angular table gives the differential outward-crossing neutron current $C_i(E)$ in zenith-angle bin $i$. It is converted to the isotropic intensity used in the optic simulation by
\begin{equation}
J_{n,\mathrm{iso}}(E)=\frac{1}{4\pi}
\sum_{i\in\mathrm{up}}\frac{C_i(E)\,\Delta\Omega_i}{\langle|\cos\theta|\rangle_i}.
\label{eq:neutronsource}
\end{equation}
Here $E$ is neutron kinetic energy, $\theta$ is the angle to the local surface normal, $\Delta\Omega_i$ is the bin solid angle and $\langle|\cos\theta|\rangle_i$ is its solid-angle-averaged normal projection. The reconstruction assumes constant intensity within each supplied $5^\circ$ angular bin.  

Primaries are launched inward from a sphere surrounding the transport geometry with the cosine-law angular distribution appropriate to isotropic intensity incident on a surface. Species and energy are importance sampled, with event weights correcting the sampling density. Independent environments are added after flux normalization. 

Each complete primary history, including all descendants, contributes the following observables. Direct charge transfer is calculated from each track's charge and whether it starts and ends inside the optic. Thus a traversing particle contributes zero net direct charge, a stopped proton contributes $+e$, and an escaping electron born in the optic contributes $+e$. Direct transfer is grouped by primary versus descendant ancestry. At optic and cavity-facing shell crossings, an additional empirical low-energy electron-emission model contributes a conditional mean and Poisson multiplicity variance. It is distinct from electrons explicitly generated and transported by Geant4. \ref{app:emission} describes the moment calculation.

Momentum transfer is the event's vector momentum balance across the optic boundaries, including momentum carried away by escaping particles; it is not restricted to particles that stop. For thermoelastic scoring, every local optic energy deposit is weighted by the signed response kernel $g(r,z)$ and summed over the complete event before forming a second moment. \ref{app:kernel} describes the spatial model.

We include a $+4$ e s$^{-1}$ allowance for residual charging processes, based on the order-unity omitted-process estimates of Ara\'{u}jo et al. \cite{araujo2005}. It is represented here as independent single-electron transfer, contributing to both the mean charging rate and the single-electron second-moment rate, rather than as an additional Geant4 mechanism. Statistical criteria target 10\% relative standard error on mean charging, 5\% on relevant second moments and 5\% changes between cumulative stages, with additional checks on signed mechanism cross terms.

\subsection{Response assumptions and numerical parameters}
\label{sec:response}
Coherent charge accumulation is computed over repeated six-hour intervals with a one-year observation time, following the finite-time treatment of \cite{shaulcoherent}. The peak envelope uses the conservative $\mathrm{sinc}$-peak prescription. Stochastic charge is driven by the physical, uncentred event second moment \cite{shaulnoise}. No charge leakage through the silica suspension or potential-dependent modification of the incident charging current is modeled. The Lorentz force uses $|\mathbf v\mathbin{\times}\mathbf B_0|=vB_0$, with adopted reference values $v=10^6$ m s$^{-1}$, $B_0=3$ nT and force-reduction factor $\eta=100$. 

The magnetic ASD prescription is $\sqrt{S_B(f)}=25.55\,\mathrm{nT\,Hz^{-1/2}}\,[f/(8.64966\times10^{-4}\,\mathrm{Hz})]^{-0.655}$. It is an adopted extrapolation inspired by solar-wind fluctuation studies such as \cite{coleman1968}, not a fitted equation quoted from that paper or a validated lunar spectrum across 0.01--1000 Hz. Under exact mean charge compensation the separate mean-charge times field-fluctuation term vanishes.

Except for the final comparison to the suspension from \cite{shapiro2025} in Figure ~\ref{fig:suspension}, mechanical forces are converted to displacement using the free-mass susceptibility $|\chi_m|=[m(2\pi f)^2]^{-1}$. This decouples the radiation- and suspension-induced responses. Torque and angular-to-longitudinal coupling are omitted. Thermoelastic deformation is instead a direct, beam-weighted surface displacement relative to the optic centre of mass. It assumes prompt thermalization and linear, homogeneous room-temperature elasticity in the adiabatic, quasistatic regime. The nominal fused-silica constants are listed in Table~\ref{tab:parameters}. Contributions within an event are added coherently; independent primary events form a Poisson process. Different final noise mechanisms are combined in quadrature without cross-spectral terms between charging, momentum and heating. 

\begin{table}[tbp]
\caption{Principal parameters, material properties and their sources. Nominal room-temperature fused-silica properties are used in the homogeneous thermoelastic model. Adopted geometry, environmental parameters and operating choices are distinguished from published detector values.}
\label{tab:parameters}
\centering\footnotesize
\setlength{\tabcolsep}{3pt}
\renewcommand{\arraystretch}{1.12}
\begin{tabular}{>{\raggedright\arraybackslash}p{.16\textwidth}>{\raggedright\arraybackslash}p{.08\textwidth}>{\raggedright\arraybackslash}p{.20\textwidth}>{\raggedright\arraybackslash}p{.22\textwidth}>{\raggedright\arraybackslash}p{.22\textwidth}}
\toprule
Parameter & Unit & This work & Source or basis & Application\\
\midrule
Mass & kg & 100 & LILA optic concept \cite{shapiro2025} & Suspended silica optic.\\
Radius; height & m & 0.23; 0.273485 & Adopted radius; derived height & Total mass and two-material densities.\\
Optic; film & -- & Silica; 6 $\mu$m mixed oxide & Silica \cite{shapiro2025}; adopted film & Two-material transport approximation.\\
Rear gap; cage clearance & mm & 5; 100 & LIGO-inspired scales \cite{ligosuspension,advancedligo2015} & Actuator and cage surrogate.\\
Electrode potentials & V & Electrode configuration (A): $400/0$; electrode configuration (B): $200/-200$ & Single-ended versus balanced \cite{balancedesd} & Compare charge coupling.\\
Reset; observation & h; d & 6; 365 & Finite-time formalism \cite{shaulcoherent} & Uncompensated coherent diagnostic.\\
Beam radius $w$ & m & 0.05 & Gaussian readout \cite{yamamoto2008} & 0.10 m beam diameter.\\
$Y$; $\nu$ & GPa; -- & 72; 0.17 & Fused silica \cite{shapiro2025,accuratussilica} & Young's modulus; Poisson ratio.\\
$\alpha$ & K$^{-1}$ & $5.5\times10^{-7}$ & Fused silica \cite{accuratussilica,degallaix2007thermal} & Linear thermal expansion.\\
$\rho$ & kg m$^{-3}$ & 2200 & Fused silica \cite{accuratussilica} & Homogeneous thermal model.\\
$c_p$ & J kg$^{-1}$ K$^{-1}$ & 740 & Fused silica \cite{accuratussilica,degallaix2007thermal} & Specific heat.\\
$v$; $B_0$; $\eta$ & m s$^{-1}$; nT; -- & $10^6$; 3; 100 & Adopted Lorentz reference parameters & Worst-axis coupling and assumed force reduction.\\
Magnetic ASD & nT Hz$^{-1/2}$ & 25.55 at\newline $8.65\times10^{-4}$ Hz;\newline slope $-0.655$ & Adopted extrapolation motivated by \cite{coleman1968} & Stochastic Lorentz force. \\
Additional charging & e s$^{-1}$ & $+4$ & Residual-process allowance based on Ara\'{u}jo et al. \cite{araujo2005} & Charging calculations.\\
Frequency interval & Hz & 0.01--1000 & Relevant midband range & All noise plots.\\
\bottomrule
\end{tabular}
\end{table}

\section{Simulation results}
\label{sec:results}
\subsection{Charging rates}
\label{sec:rates}
\begin{figure}[tbp]
\centering\includegraphics[width=\textwidth]{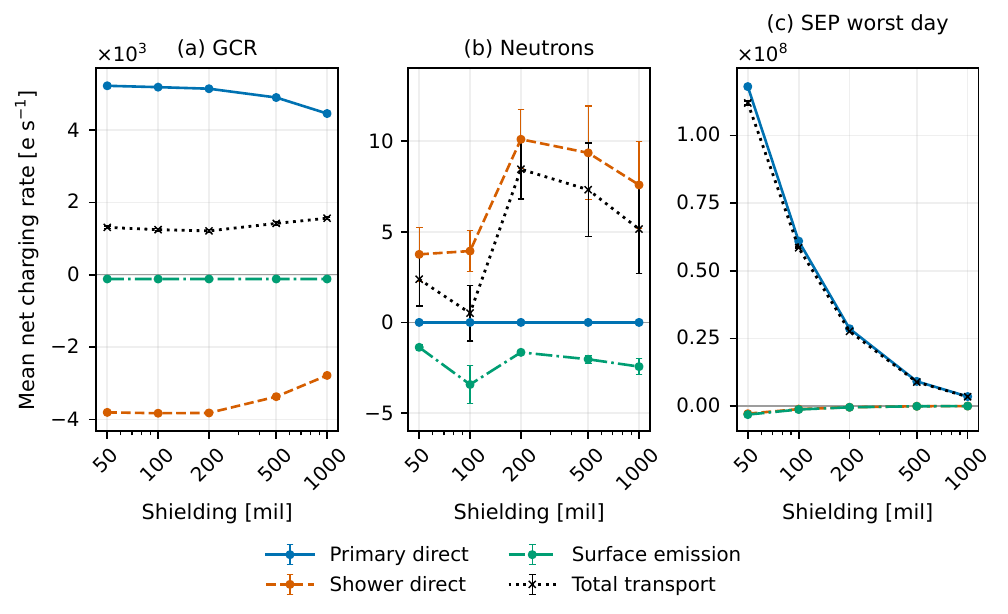}
\caption{Signed mean charging rates versus aluminum-equivalent shielding thickness for (a) solar-minimum GCRs, (b) lunar albedo neutrons and (c) the worst-day SEP component. Primary and shower direct transfer are distinguished from additional low-energy surface emission. The total is their signed sum; error bars are Monte Carlo standard errors, including mechanism covariance for the total.}
\label{fig:rates}
\end{figure}

Figure~\ref{fig:rates} separates primary direct transfer, shower direct transfer, and the additional surface-emission contribution. At 50 mil, the GCR case has primary direct charging of $\valGcrPrimaryFifty$ e s$^{-1}$, shower direct charging of $\valGcrShowerFifty$ e s$^{-1}$, and modeled surface emission of $\valGcrEmissionFifty$ e s$^{-1}$. The resulting GCR net rate is $\valGcrMeanFifty$ e s$^{-1}$. The neutron contribution is $\valNeutronMeanFifty$ e s$^{-1}$, giving a combined background rate of $\valBackgroundMeanFifty$ e s$^{-1}$ before the independent allowance. Shielding has a negligible effect on the background charging rates.

The SEP mean rate decreases from $\valSepMeanFifty$ to $\valSepMeanThousand$ e s$^{-1}$ between 50 and 1000 mil, a factor of about \valSepShieldFactor{}. Direct primary charging dominates this mean; shower and surface terms reduce it. Shielding modifies secondary production and cancellation as well as primary transmission. Consequently a signed net charging rate need not decrease monotonically with material thickness, and its variation cannot be inferred from attenuation alone.

\subsection{Coherent charging and continuous compensation}
\label{sec:coherent}
Without continuous compensation, $Q(t)=\dot Q t$ between zero-charge resets. Following \cite{shaulcoherent}, the force gives terms proportional to $t$, $t^2$ and $t$ from linear Coulomb coupling, quadratic Coulomb coupling and Lorentz coupling, respectively. Write $\psi=A\dot Q/m$, $\xi=B\dot Q^2/m$, and $\phi=\dot Q vB_0/(m\eta)$, where $\eta$ is the magnetic reduction factor. For observation time $T$, reset period $\tau$ and harmonic index $n$, the peak-envelope calculation used here is
\begin{equation}
H_x(f)=\frac{\sqrt{2/T}}{(2\pi f)^2}
\max_{\pm}\left|\frac{\tau T}{2\pi n}
\left[\frac{\xi\tau}{\pi n}+\mathrm i(\psi+\xi\tau\pm\phi)\right]\right|,
\label{eq:coherent}
\end{equation}
Here $n=\max\{1,\mathrm{round}(f\tau)\}$. The sign maximization accounts for the unspecified Lorentz-axis orientation while preserving the common charge phase. The Coulomb terms are combined as complex amplitudes before taking a magnitude; they are not independent noise powers. Replacing the finite-time line-shape factor by its peak produces the conservative envelope, rather than a literal broadband spectrum at every frequency.

\begin{figure}[tbp]
\centering\includegraphics[width=\textwidth]{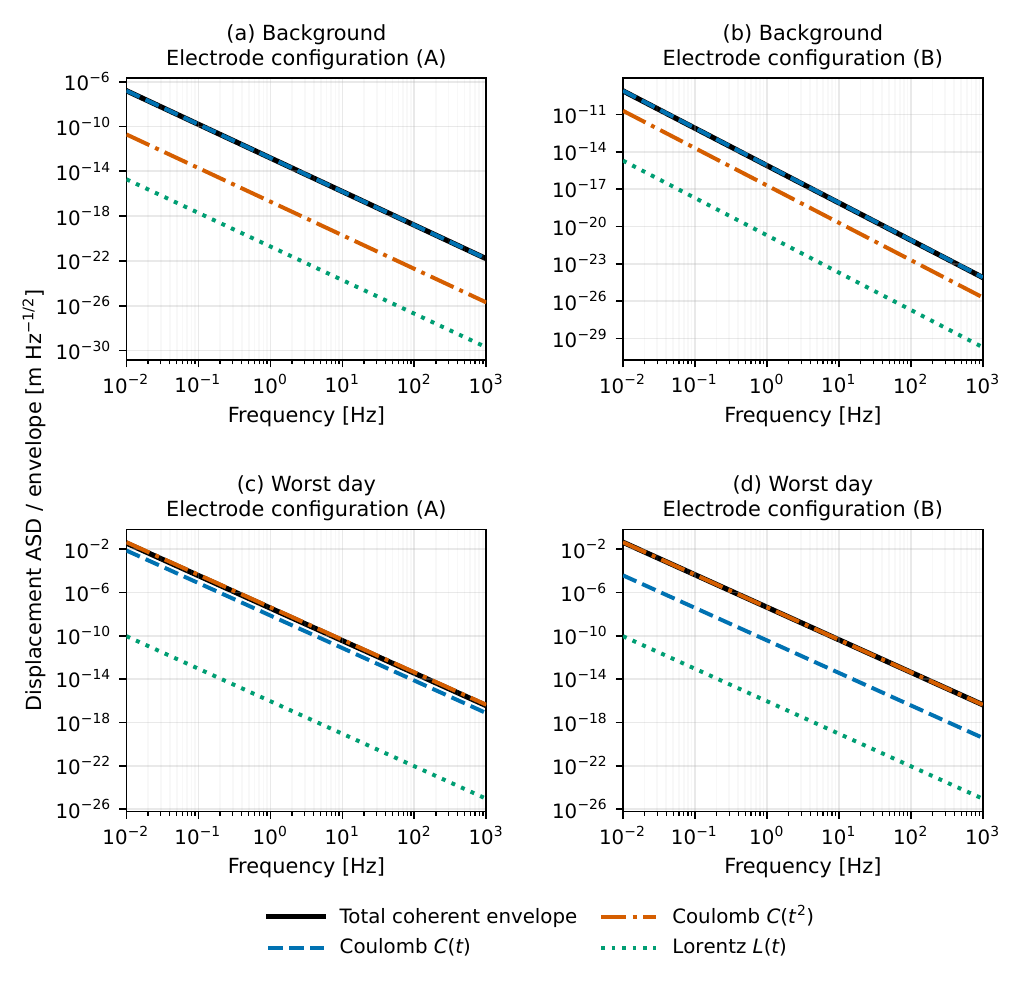}
\caption{Coherent charging displacement envelopes for 100 mil shielding without continuous compensation: (a) electrode configuration (A), background; (b) electrode configuration (B), background; (c) electrode configuration (A), storm; (d) electrode configuration (B), storm. Linear Coulomb $C(t)$, quadratic Coulomb $C(t^2)$, Lorentz $L(t)$ and their phase-consistent total are shown. These are conservative harmonic-peak envelopes (sinc term set to 1).}
\label{fig:coherent}
\end{figure}

Figure~\ref{fig:coherent} shows the 100 mil cases without continuous discharge. For background, $\bar R=\valBackgroundRateHundred$ e s$^{-1}$ including the allowance, and the linear Coulomb term dominates. Electrode configuration (B) strongly reduces it. For the storm, the mean is $\valStormRateHundred$ e s$^{-1}$ and the quadratic term dominates because it scales as $\dot Q^2$. The similar values of $B$ in the two electrode configurations then limit the benefit of balanced drive alone. These curves are diagnostics of the assumed no-compensation trajectory. The choice of electrode configuration thus has little effect on the worst-day coherent charging noise envelope when compared to the background case. 

\begin{figure}[tbp]
\centering\includegraphics[width=\textwidth]{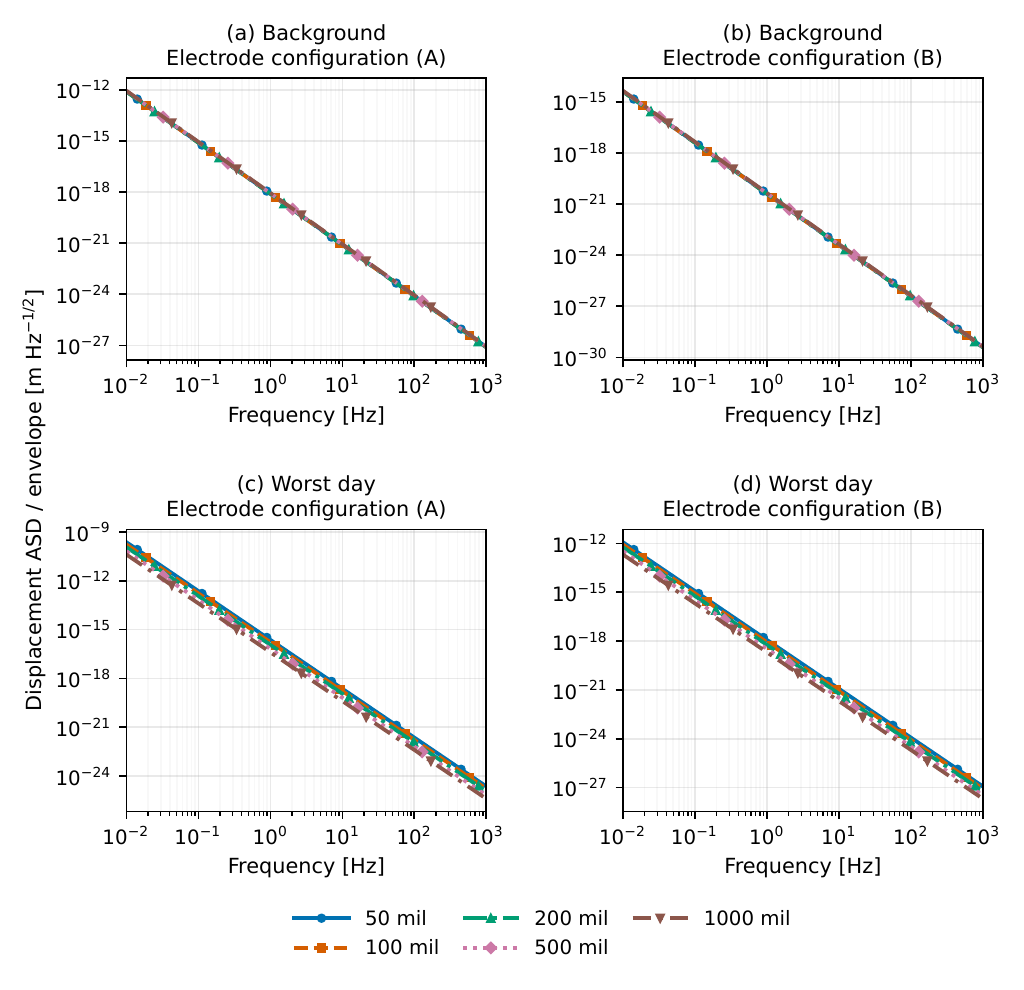}
\caption{Discharge shot-noise displacement ASD alone for 50, 100, 200, 500 and 1000 mil shielding under exact continuous mean compensation: (a) electrode configuration (A), background; (b) electrode configuration (B), background; (c) electrode configuration (A), worst-day storm; (d) electrode configuration (B), worst-day storm. Independent single-electron transfers use the same uniform-charge force coupling as radiation charging.}
\label{fig:discharge}
\end{figure}

Exact mean compensation is adopted following calculation of the optimization integral in Equation~(\ref{eq:optimization}). At 100 mil, the required reference rates are approximately $\valBackgroundRateHundred$ e s$^{-1}$ for background and $\valStormRateHundred$ e s$^{-1}$ during the prescribed storm, corresponding to about $\valBackgroundCurrentHundred$ and $\valStormCurrentHundred$ A. With exact mean cancellation and zero initial charge, Equation~(\ref{eq:coherent}) vanishes. The displacement noise induced by discharge currents is shown in Figure~\ref{fig:discharge}, calculated with  Equation \ref{eq:dischargeshot} for each shielding thickness. Shielding has little impact in practice, as the change in charging rates is linear even for SEP.  

\subsection{Stochastic charging noise}
\label{sec:chargenoise}
Let $n=q/e$ be the signed number of elementary charges transferred by one event, and let $\lambda_r=\sum_i R_i\langle n_i^2\rangle$ be its physical uncentred second-moment rate, including the independent allowance. Thus $\lambda_r$ has units s$^{-1}$; rates quoted in e$^2$ s$^{-1}$ use elementary-charge counting units. The radiation-current PSD is $S_{I,r}=2e^2\lambda_r$. The charge random-walk spectrum and longitudinal displacement spectrum are
\begin{eqnarray}
S_{Q,r}(f)&=&\frac{2e^2\lambda_r}{(2\pi f)^2},\nonumber\\
S_{x,r}(f)&=&|\chi_m(f)|^2
\left[k_r^2 S_{Q,r}(f)+\left(\frac{\bar Q v}{\eta}\right)^2 S_B(f)\right],
\label{eq:chargenoise}
\end{eqnarray}
Here $k_r=\max_{Q\in\mathcal C}|A+2BQ|+vB_0/\eta$, where $\mathcal C=\{0,\dot Q\tau\}$ contains the endpoints of the mean reset trajectory. Checking both electrostatic slopes prevents accidental cancellation at one cycle endpoint from suppressing the reported noise. Under exact continuous compensation $\mathcal C=\{0\}$. The magnetic-field term uses the largest mean charge magnitude $\bar Q$ over that cycle, and $S_B$ is the prescribed field-fluctuation spectrum. 

As a diagnostic comparison with the uncompensated trajectory, at 100 mil and 0.1 Hz compensation reduces the storm radiation charging-noise contribution by factors of $\valCompensationReductionA$ in electrode configuration (A) and $\valCompensationReductionB$ in electrode configuration (B). This is the removal of the large accumulated-charge contribution to $A+2B\bar Q$, not suppression of the incident current fluctuations. Once compensated, electrode configuration (A) has \valForceRatio{} times the radiation charging ASD of electrode configuration (B), reflecting the ratio of the linear electrostatic couplings.

\begin{figure}[tbp]
\centering\includegraphics[width=\textwidth]{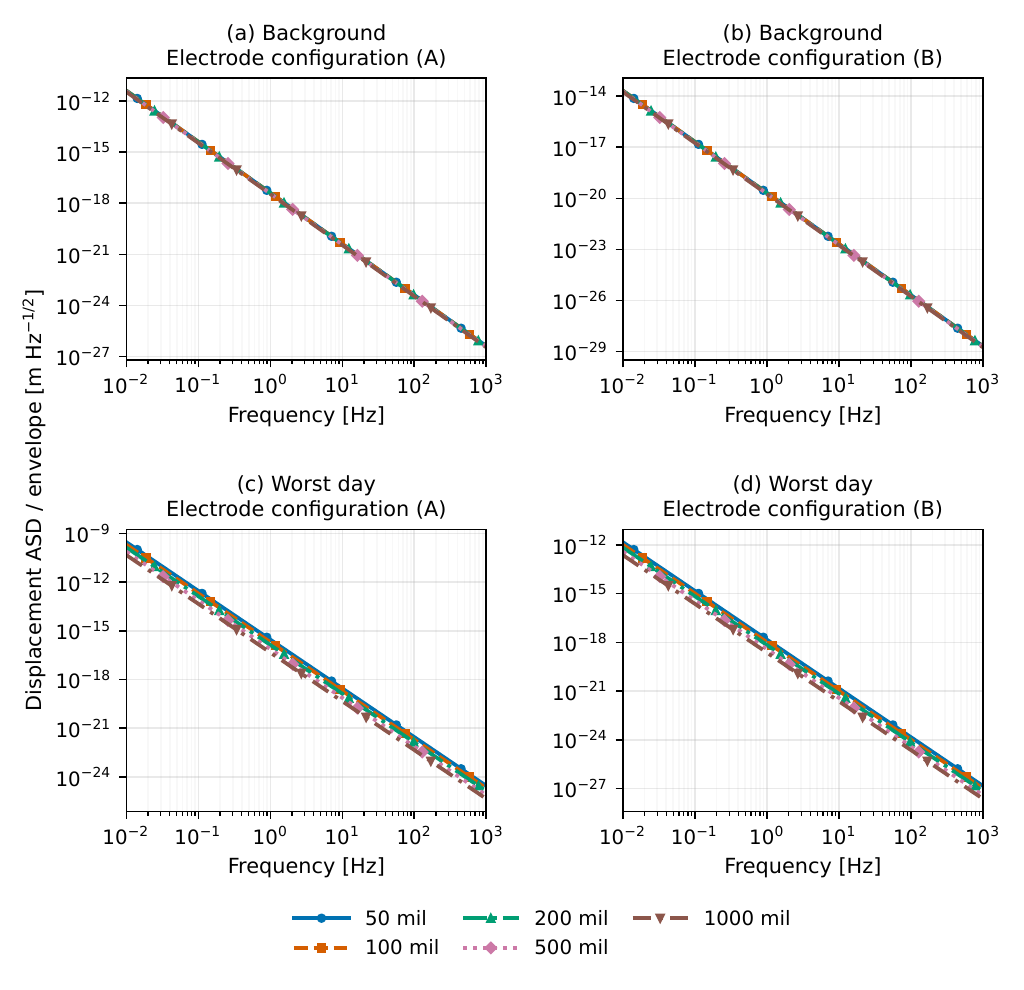}
\caption{Mean-compensated radiation charging-noise displacement ASD for 50, 100, 200, 500 and 1000 mil shielding: (a) electrode configuration (A), background; (b) electrode configuration (B), background; (c) electrode configuration (A), worst-day storm; (d) electrode configuration (B), worst-day storm. Exact continuous mean balance maintains zero mean charge without changing the radiation-induced stochastic current. Neither coherent charging nor discharge-source shot noise is included.}
\label{fig:chargenoise}
\end{figure}

\subsection{Momentum-kick noise}
\begin{figure}[tbp]
\centering\includegraphics[width=\textwidth]{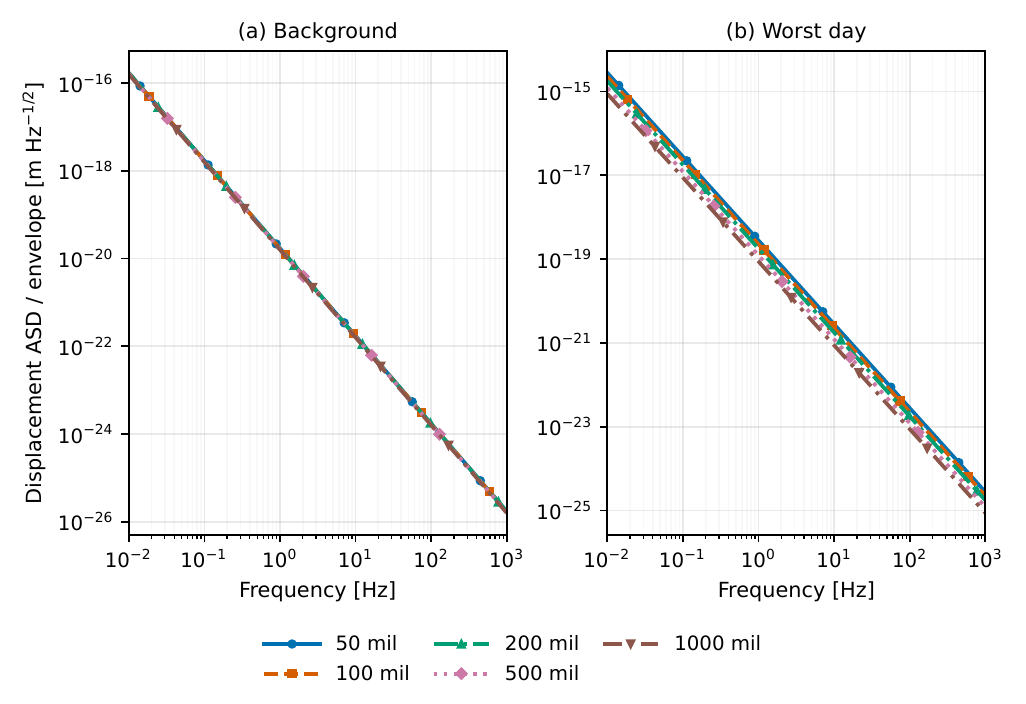}
\caption{Momentum-kick displacement ASD for 50, 100, 200, 500 and 1000 mil shielding in (a) background and (b) worst-day storm conditions. The same 100 kg free-mass response and isotropic one-axis impulse partition are used for both environments. Electrode configuration does not change this contribution.}
\label{fig:kick}
\end{figure}

The net event impulse $\mathbf J$ includes momentum delivered by incident and secondary particles and momentum carried away by escaping particles. The current analysis retains its magnitude and assumes an isotropic longitudinal partition, $\langle J_z^2\rangle=\langle J^2\rangle/3$, as a reference approximation. For event-rate bins $R_j$ with impulse magnitude $J_j$, the one-sided displacement PSD is
\begin{equation}
S_{x,\mathrm{kick}}(f)=\frac{2}{3m^2(2\pi f)^4}\sum_j R_j J_j^2.
\label{eq:kick}
\end{equation}
This follows the independent-impulse shot-noise treatment discussed for space-based GW detectors \cite{lisa1998}. The resulting acceleration ASD is white and the free-mass displacement ASD scales as $f^{-2}$. Momentum kick noise does not depend on the electrode bias or continuous charge compensation. Figure \ref{fig:kick} shows the displacement ASD for kick noise. At 100 mil it is $\valKickBackgroundOneHz$ $\asdunit$ at 1 Hz for background and approximately $\valKickStormOneHz$ $\asdunit$ for the full storm. Equation~(\ref{eq:kick}) omits impulse anisotropy relative to the cylindrical optic, torque, and the actual suspension response.

\subsection{Thermoelastic displacement}
\label{sec:thermoelastic}
Local heat deposited by a radiation event produces elastic deformation that is sampled by the optical beam. The signed event displacement amplitude is $a_h=\sum_{s\in h}g(r_s,z_s)\Delta E_s$, with energy in joules. In the adiabatic, quasistatic approximation the Fourier response is $a_h/(\mathrm i2\pi f)$ and the one-sided PSD is $S_{x,\mathrm{TE}}=2\mathcal R\langle a_h^2\rangle/(2\pi f)^2$. Here $\mathcal R\langle a_h^2\rangle$ denotes the flux-normalized event second-moment rate. Deposits from the same event are summed with their signs before squaring, so secondary-shower correlations are retained.

Unlike a rigid-body force, this contribution is already a displacement relative to the optic centre of mass. It must not be passed through a pendulum susceptibility a second time. The spatial kernel removes the need to approximate a stopping particle by a constant stopping power extending through the optic. The calculation retains the linear thermoelastic and independent-event assumptions of the treatment in \cite{yamamoto2008}, with the kernel and numerical checks described in \ref{app:kernel}.

\begin{figure}[tbp]
\centering\includegraphics[width=\textwidth]{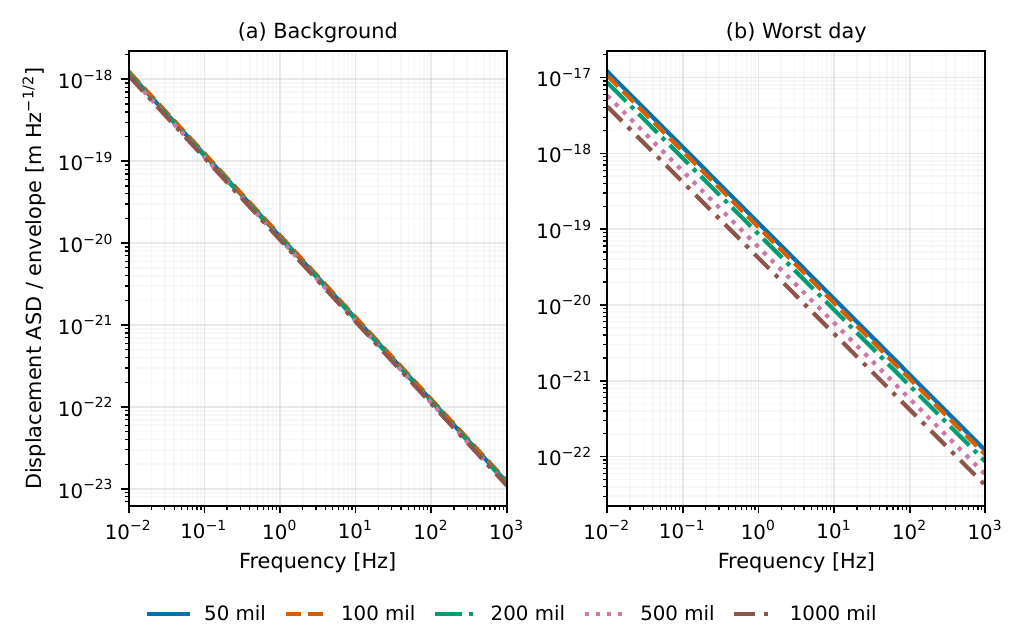}
\caption{Radiation-induced thermoelastic displacement ASD for the indicated aluminum-equivalent shielding thicknesses in background and storm conditions. The response is the Gaussian-beam-weighted front-face displacement relative to the optic centre of mass. The plotted 0.01--1000 Hz band uses the spatial adiabatic, quasistatic kernel.}
\label{fig:te}
\end{figure}

Figure~\ref{fig:te} shows the $f^{-1}$ ASD dependence of the adopted adiabatic response. At 100 mil and 1 Hz, the values are $\valTeBackgroundOneHz$ $\asdunit$ for background and $\valTeStormOneHz$ $\asdunit$ for the full storm. The result is a direct deformation estimate, not a prediction of total optical phase noise: coating-specific thermoelasticity and thermorefractive effects are not included. 

\section{Other radiation effects outside the displacement budget}
\label{sec:other}
The preceding results describe direct radiation-driven mechanical and thermoelastic responses of the optic. Radiation damage and transient effects in optoelectronics require additional consideration. We give screening estimates rather than include unqualified electronics-noise terms in the mechanical budget. Mitigation is possible in many cases.

\subsection{Accumulated dose and displacement damage}
\label{sec:dose}
The mission screening calculation supplied for this study assumes a notional Earth-departure phase lasting 25 h in an orbit with 185 km perigee altitude, 71\,656 km apogee altitude and inclination $28.5^\circ$, followed by a three-day translunar coast and 30 years on the lunar surface. OMERE v5.8 was used with AE9/AP9 trapped-particle models and ESP solar-proton models, employing SHIELDOSE-2 spherical-aluminum dose transport and the Summers displacement-damage treatment \cite{omere,ae9ap9,esp,esp2000,shieldose2,summers1993}. Neutron-induced dose is neglected. At a 95\% confidence level, the departure and coast phases contribute 8741 and 170 rad(Si), with DDDs $5.405\times10^7$ and $1.988\times10^6$ MeV g$^{-1}$. Lunar operation contributes 21\,250 rad(Si) and $2.485\times10^8$ MeV g$^{-1}$ per year. Adding the transit terms and thirty annual values gives 646 krad(Si) and $7.51 \times10^9$ MeV g$^{-1}$. This number is extremely conservative as it assumes a 95\% confidence level bound year each year, as opposed to a 95\% bound for the full 30 year period. These doses warrant radiation-hardness assurance for precision electronics. Part selection, process qualification, shielding, bias and temperature control, calibration margin and circuit-level recovery are established elements of that process \cite{esahandbook,ecssrha}. 

Radiation-induced changes in laser threshold current, efficiency and wavelength depend on technology \cite{johnston2003,johnston2004}. Proton-induced redshifts have been reported for particular AlGaInAs/InP semiconductor lasers \cite{bryant2026}. The response of a LIGO-like diode-pumped Nd:YAG laser \cite{ligolaser} is unknown in these radiation environments - shifts in laser power and frequency must be investigated. Laser power control and frequency stabilization provide possible mitigation paths.

For an illustrative silicon photodetector, displacement-induced leakage current may be estimated using \cite{moll1999,neutronconversion}
\begin{equation}
\Delta I_{\mathrm{dark}}=\alpha_I\Phi_{\mathrm{eq}}V_{\mathrm{Si}},\qquad
\Phi_{\mathrm{eq}}=\frac{D_{\mathrm{DDD}}}{\mathrm{NIEL}_{n,1\,\mathrm{MeV}}}.
\label{eq:dark}
\end{equation}
Here $\alpha_I$ is the leakage-current damage coefficient, $\Phi_{\mathrm{eq}}$ is the 1 MeV neutron-equivalent fluence in silicon, and $V_{\mathrm{Si}}$ is the sensitive volume. Taking $\mathrm{NIEL}_{n,1\,\mathrm{MeV}}=2.04\times10^{-3}$ MeV cm$^2$ g$^{-1}$, $\alpha_I=4\times10^{-17}$ A cm$^{-1}$, and a 1 mm$^2$ by 100 $\mu$m collection volume gives $\Phi_{\mathrm{eq}}=3.68\times10^{12}$ cm$^{-2}$ and $\Delta I_{\mathrm{dark}}=14.7$ nA \cite{neutronconversion,moll1999}. The coefficient refers to a specified annealing convention and approximately $20\,^{\circ}$C; actual conditions may vary on the lunar surface.

The associated shot noise is $\sqrt{2e\Delta I_{\mathrm{dark}}}=68.7$ fA Hz$^{-1/2}$. For an optical operating point of 10 mW and assumed responsivity 0.8 A W$^{-1}$, the photocurrent is 8 mA and optical shot noise is 50.6 pA Hz$^{-1/2}$ - far higher than the dark current shot noise. LISA photoreceiver testing provides relevant evidence for device qualification: after exposures up to 237 krad and $10^{12}$ protons cm$^{-2}$ equivalent, almost all tested quadrant photodiodes met the study's validation criteria despite measured changes in properties \cite{lisaphotodiodes}. 

\subsection{Single-event effects in optics and electronics}
Single-event effects can produce digital upsets, analog transients, destructive failures, or interfere with sensor operation. Shrinking device feature sizes and rail voltages have increased susceptibility to single event effects in general, with techniques such as near threshold computing at particular risk  \cite{fleetwood2021postmoore, trippe2015electron}. Destructive effects such as burnout, latchup, and stuck bits threaten mission success ~\cite{sexton2003destructive,ball2021finfetlatchup,trippe2025sedd}. These effects can conceivably impact the measurement in a number of ways, such as introducing noise in digitizers, disrupting calculations, affecting power stability, or creating sticky points in actuators. 

While part-specific analysis is outside the scope of this paper, methods exist to minimize the introduction of component-level radiation technical noise. Parts can be selected based on flight heritage or ground test data \cite{label1996see}. Error correction and memory scrubbing, redundancy and voting, watchdog recovery, and current isolation are possible digital and system-level mitigations. Analog mitigation may use appropriate bandwidth, filtering, clamps, transient recovery and independent monitoring, but susceptibility is application dependent. These effects are excluded from the mechanical displacement curves pending device selection and circuit tests.

One particularly important case is an ionizing strike to the photodiode. Charge collected from the ion can directly enter the optical readout. We use GCR data from the RADOM photodetector from Chandrayaan-1 as an illustrative case ~\cite{DACHEV2011779}, noting that no photodiode configuration or noise requirements are set for the GDC. These diodes are 200 mm$^2$, 300 $\mu$m thick silicon behind about 66 mil (1.68 mm) of Al equivalent shielding. At room temperature the pair-creation energy is about 3.62 eV, giving $k_E=4.43\times10^{-14}$ C MeV$^{-1}$. For independent short charge impulses, the low-frequency one-sided current PSD is
\begin{equation*}
\begin{array}{l}
S_I(f)=2k_E^2\sum_i R_i\langle E_{\mathrm{dep},i}^{2}\rangle\\[3pt]
\end{array}
\end{equation*}

where $R_i$ is the observed rate in the $E_i$ bin from ~\cite{DACHEV2011779}. Summing over all bins gives $\sum_i R_i\langle E_{\mathrm{dep},i}^{2}\rangle=9.43$ MeV$^2$ s$^{-1}$ and $S_I(f)=3.70\times10^{-26}$ A$^2$ Hz$^{-1}$. 

To illustrate conversion to displacement, the historical Hanford O2 sensing gain $K_0=3.834\times10^9$ A m$^{-1}$ and coupled-cavity pole $f_{\mathrm{cc}}=360$ Hz are available from interferometer calibration \cite{ligocalibration2017}. The simplified single-pole benchmark gives
\begin{eqnarray*}
\sqrt{S_{x,\mathrm{SET}}(f)}&=&\frac{\sqrt{S_I(f)}}{K_0}\sqrt{1+(f/f_{\mathrm{cc}})^2}
\end{eqnarray*}
At 0.1 Hz and 100 Hz the estimates are $5.02\times10^{-23}$ and $5.20\times10^{-23}$ $\asdunit$, respectively, smaller than the noise envelope maximums of $2.14\times10^{-17}$ and $1.24\times10^{-22}$ $\asdunit$ respectively.

These results are generally conservative due to the low shielding and large diode area. This suggests diode transients induced by radiation are unlikely to dominate the radiation noise on the LILA detectors. The final chosen detectors need to be analyzed to fully estimate this effect in a realistic GDC implementation to see if mitigations are required. 

\subsection{Charge coupling to electrode noise}
For electrode voltages $V_j$, the force generally depends on both charge and voltage. Small fluctuations contain 
\begin{equation}
    \delta F_V =
\sum_j \frac{\partial F_0}{\partial V_j}\delta V_j
+\overline{Q}\sum_j\frac{\partial A}{\partial V_j}\delta V_j
+\delta Q\sum_j\frac{\partial A}{\partial V_j}\delta V_j
\end{equation} where $\delta F_V$ is the voltage contribution to the noise and $\delta V_j$ is the voltage at the jth electrode. 

The first term is the electrode noise mixing. It is charge independent and therefore unaffected by radiation, making it outside the scope of this paper. The second term disappears under continuous discharging. Under continuous charging, it would scale according to

\begin{equation}
\sqrt{S_{x,2}(f)}=\frac{|\dot{Q}|\tau c_{\mathrm{eff}}\sqrt{S_V(f)}}{m(2\pi f)^2},\qquad c_{\mathrm{eff}}=8.55~\mathrm{NC^{-1}V^{-1}}
\end{equation}
where $\sqrt{S_V(f)}$ must be assumed. As an illustrative example, to match 10\% of the 0.1 Hz coherent ASD with 6 hour discharge periods, the electrodes would have to have a $\sqrt{S_V(f)}=0.825~\mathrm{V} \mathrm{Hz^{-1/2}}$, well outside the expected operating budget. For example, LIGO only allows $\sqrt{S_V(f)}=33~\mathrm{nV} \mathrm{Hz^{-1/2}}$ at 400 V. Given the current assumption is continuous discharging (when this term goes to zero) and even in the worst case the noise required on the electrodes for charge-voltage coupling to become apparent would already have exceeded reasonable limits, this second term is neglected.

The remaining charge--voltage contribution is the second-order mixing term $\delta F_{QV}(t)=\delta Q(t)\sum_j c_j\delta V_j(t)$, where $c_j=\partial A/\partial V_j$.  For independent, equal voltage noise on the bias and signal electrodes, the FEMM derivatives give $\sqrt{S_A(f)}=8.55\sqrt{S_V(f)}$.  Because this term is a product of two stochastic processes, its exact force PSD is the convolution $S_{F,QV}(f)=\int_{-\infty}^{\infty}S_Q(\nu)S_A(f-\nu)\,d\nu$; therefore, it cannot be bounded from the voltage ASD at a single frequency.  As a screening estimate, defining the integrated voltage noise by $V_{\mathrm{rms}}^2=\int S_V(f)\,df$ gives $\sqrt{S_{x,QV}}\simeq(8.55V_{\mathrm{rms}}/k_Q)\sqrt{S_{x,Q}}$.  At $0.01~\mathrm{Hz}$ for the 100-mil, configuration-B uncompensated cases, this becomes $1.45\times10^{-14}V_{\mathrm{rms}}~\mathrm{m}/\sqrt{\mathrm{Hz}}$ for the background and $7.11\times10^{-13}V_{\mathrm{rms}}~\mathrm{m}/\sqrt{\mathrm{Hz}}$ for the worst-day environment, with $V_{\mathrm{rms}}$ expressed in volts.  For $V_{\mathrm{rms}}=1~\mathrm{mV}$, these contributions are only $0.069\%$ and $3.0\times10^{-5}\%$ of the corresponding charging ASDs.  They would reach $10\%$ of the charging ASD only for approximately $V_{\mathrm{rms}}=0.145~\mathrm{V}$ and $331~\mathrm{V}$, respectively. Thus, the mixing term is expected to be negligible for millivolt-level integrated electrode noise, although a rigorous bound requires the measured voltage-noise spectra, bandwidths, and inter-electrode correlations. 

While these results motivate leaving these terms out of the overall noise curve, final analysis of the charge-voltage coupling of the final electrode design is required. Also neglected are other possible charge-voltage couplings, such as from power supplies, which must be accounted for in the final analysis. 

\subsection{Nonuniform charge and charge migration}
The optical coating and substrate are insulating. Localized charge and field-driven redistribution are physically credible on silica test masses \cite{chargepatch,balancedesd}. This study nevertheless retains the uniform-charge and uniform-discharge approximation. Noise from uneven charging or discharging from surface defects and insulating behavior will be studied when the coatings are finalized. 

\subsection{Thermoelastic displacement on non-TM mirrors}
Advanced LIGO has many mirrors smaller than the test masses, and which have smaller incident radii \cite{advancedligo2015}. These may introduce more thermoelastic noise as $g_w \propto 1/w^2$. Additional mirrors are neglected here, as this would require a finalized end stage to understand the rate, severity, and coupling to the final optical readout. 
\section{Implications for the LILA Observatory}
\label{sec:implications}
\subsection{Lunar Deformation Channel}
The LDC does not use the suspended free-mass readout modeled here. Direct thermoelastic surface motion remains relevant, whereas rigid-body charging and kick forces cannot simply be converted using the GDC free-mass susceptibility. At 50 mil the full-storm thermoelastic ASD is $\valTeStormFiftyOneHz$ $\asdunit$ at 1 Hz. Formally extending the adiabatic $f^{-1}$ amplitude dependence to 1 mHz gives $\valTeStormFiftyMilliHz$ $\asdunit$, below a reference displacement sensitivity of $10^{-15}$ $\asdunit$ \cite{creighton2025}. This estimate must be refined when the mirror coatings on the LDC are decided.

\subsection{Geodesic Deformation Channel}
\label{sec:gdc}
\begin{figure}[tbp]
\centering\includegraphics[width=\textwidth]{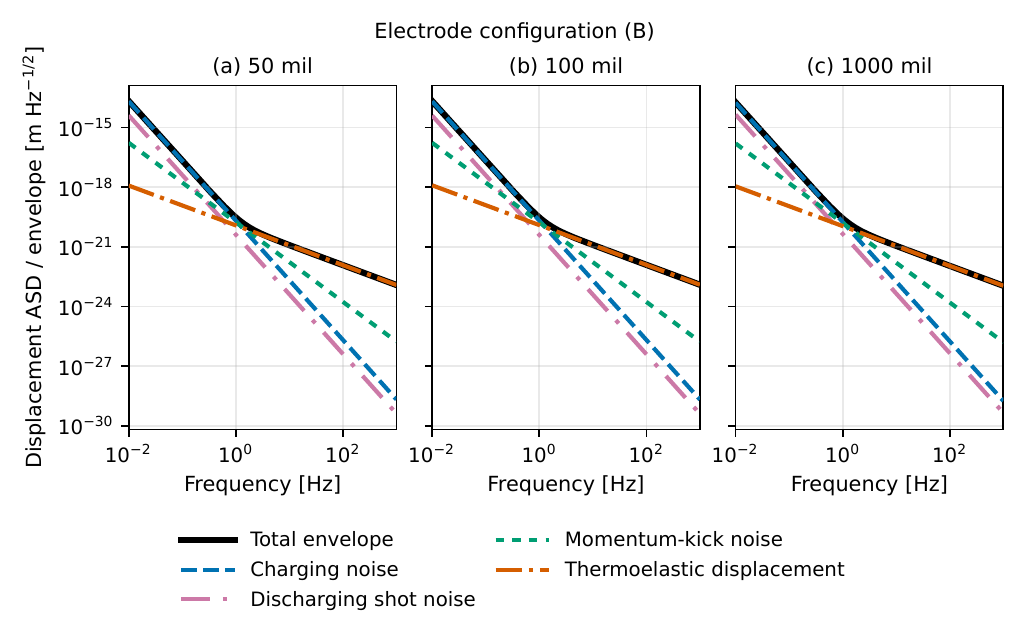}
\caption{Contributions to the total modeled displacement ASD envelope for the background environment and electrode configuration (B), with (a) 50 mil, (b) 100 mil and (c) 1000 mil shielding. Discharge shot noise, radiation charging noise, momentum-kick noise and direct thermoelastic displacement are shown. Exact continuous mean balance makes the charging-coherent term zero. Mechanical contributions use a free 100 kg mass.}
\label{fig:bg_budget}
\end{figure}

\begin{figure}[tbp]
\centering\includegraphics[width=\textwidth]{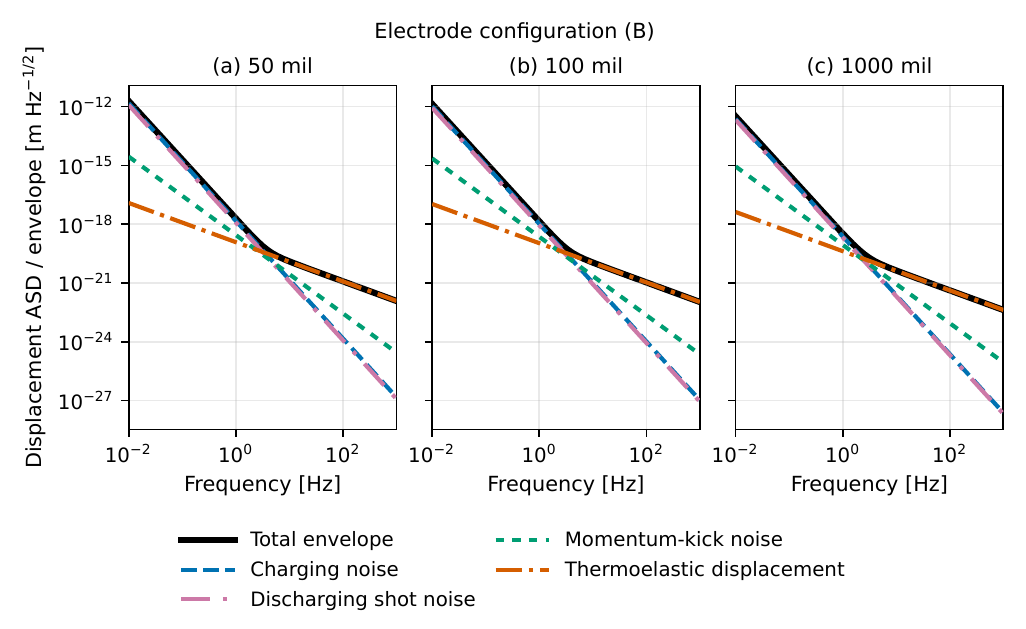}
\caption{The same budget as Figure~\ref{fig:bg_budget} for the worst-day storm scenario, comprising the prescribed SEP spectrum plus solar-minimum GCRs and lunar neutrons. Electrode configuration (B) and exact continuous mean compensation are used for (a) 50 mil, (b) 100 mil and (c) 1000 mil shielding.}
\label{fig:storm_budget}
\end{figure}

Figures~\ref{fig:bg_budget} and \ref{fig:storm_budget} show the complete modeled mechanical and thermoelastic budgets for 50, 100 and 1000 mil shielding using electrode configuration (B) and continuous mean compensation. Radiation charging, independent discharge shot noise, momentum kicks and direct thermoelastic deformation are retained. Their total is the quadrature envelope used for design comparison, not an assertion that an uncompensated deterministic signal is stationary noise.

For background, radiation charging noise dominates the low-frequency response. At 100 mil and 0.1 Hz, the total is $\valBackgroundTotalTenthHz$ $\asdunit$, with radiation charging $\valBackgroundChargingTenthHz$, discharge shot noise $\valBackgroundDischargeTenthHz$, and momentum kicks $\valBackgroundKickTenthHz$ $\asdunit$. At 1 Hz the total is $\valBackgroundTotalOneHz$ and at 100 Hz it is $1.24\times10^{-22}$ $\asdunit$. The relative importance of momentum transfer increases with frequency before direct thermoelastic deformation controls the upper-frequency response. The separate signed charge rates and secondary yields can vary with shielding even when their combined noise changes weakly.

During the prescribed storm, charging and discharge shot noise are comparable at low frequency. At 100 mil and 0.1 Hz they are approximately $\valStormChargingTenthHz$ and $\valStormDischargeTenthHz$ $\asdunit$, with total $\valStormTotalTenthHz$ $\asdunit$. The total falls to $\valStormTotalOneHz$ $\asdunit$ at 1 Hz and $\valStormTotalHundredHz$ $\asdunit$ at 100 Hz, where thermoelastic deformation dominates. Increasing shielding from 50 to 1000 mil reduces the 0.1 Hz storm total from $\valStormTotalFiftyTenthHz$ to $\valStormTotalThousandTenthHz$ $\asdunit$, a factor of \valStormShieldFactor{}. This is smaller than the corresponding reduction in mean SEP charging because stochastic noise depends on event second moments and on the discharge rate through a square root.

\begin{figure}[tbp]
\centering\includegraphics[width=\textwidth]{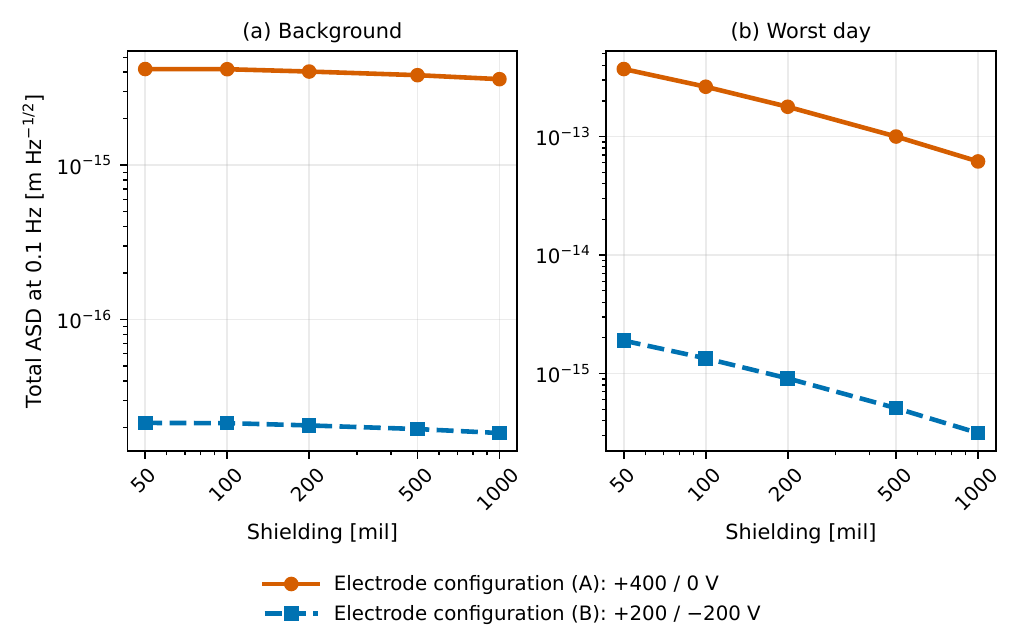}
\caption{Effect of electrode configuration on total modeled displacement ASD at 0.1 Hz versus aluminum-equivalent shielding thickness for (a) background and (b) worst-day storm conditions. Electrode configuration (A) uses $400/0$ V and electrode configuration (B) uses $200/-200$ V. Both use exact continuous mean balance.}
\label{fig:electrode}
\end{figure}

Figure~\ref{fig:electrode} compares the electrode configurations across all five thicknesses at 0.1 Hz, where charge-related terms dominate. At 100 mil and under exact compensation, the total ASD for electrode configuration (A) is \valBackgroundDriveRatio{} times that for electrode configuration (B) in background and \valStormDriveRatio{} times that for electrode configuration (B) in the storm. Shielding is the stronger lever on storm exposure, while electrode symmetry chiefly controls the conversion of residual stochastic charge into force.

\begin{figure}[tbp]
    \centering
    \captionsetup[subfigure]{skip=2pt}

    \begin{subfigure}{0.75\textwidth}
        \centering
        \makebox[\linewidth][c]{%
    \hspace*{-4mm}%
    \includegraphics[
        width=\linewidth,
        trim={0 20mm 0 0},
        clip
    ]{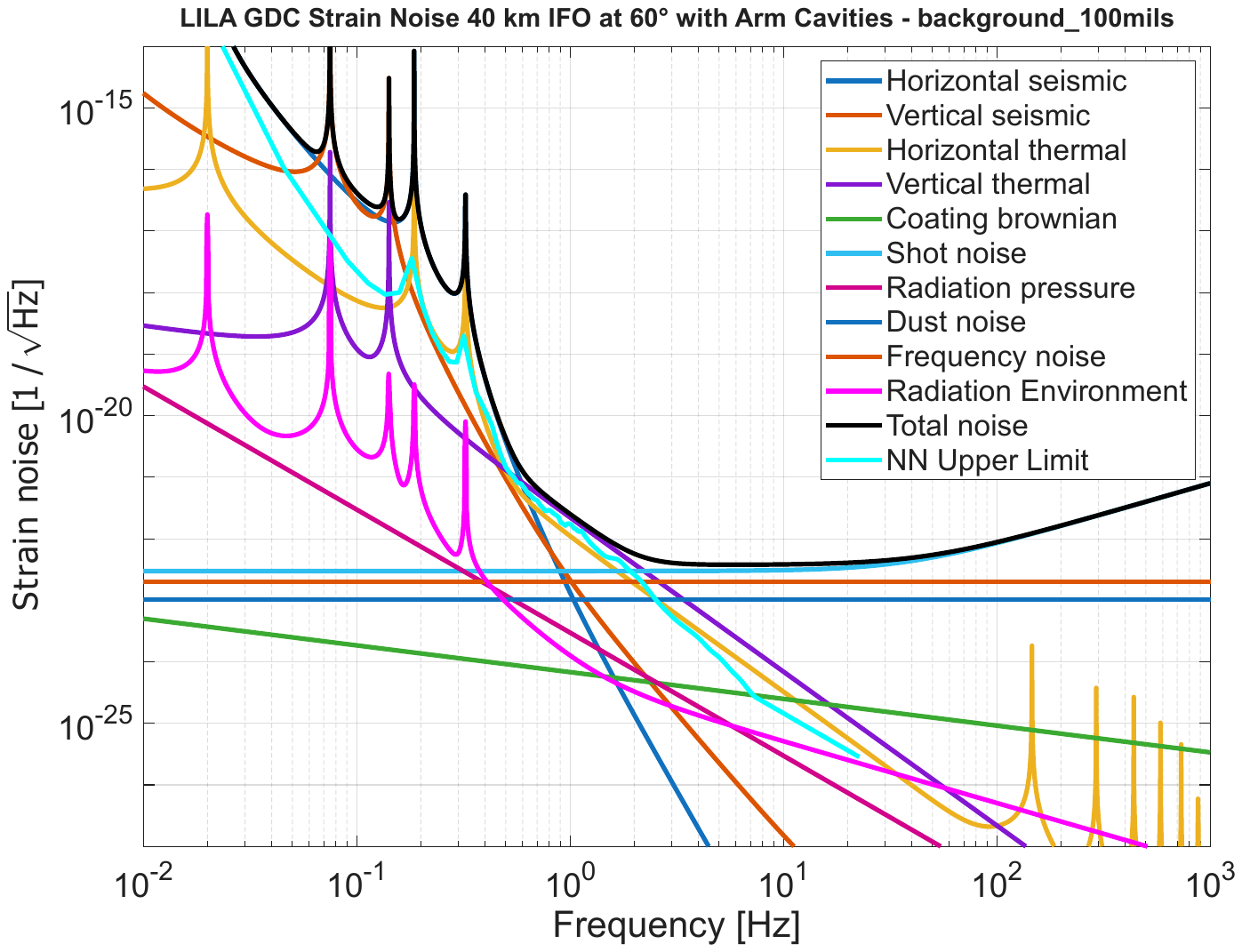}%
    \hspace*{4mm}%
}
        \caption{Background}
        \label{fig:suspension-background}
    \end{subfigure}

    \par

    \begin{subfigure}{0.75\textwidth}
        \centering
        \makebox[\linewidth][c]{%
    \hspace*{-4mm}%
    \includegraphics[
        width=\linewidth,
        trim={0 20mm 0 0},
        clip
    ]{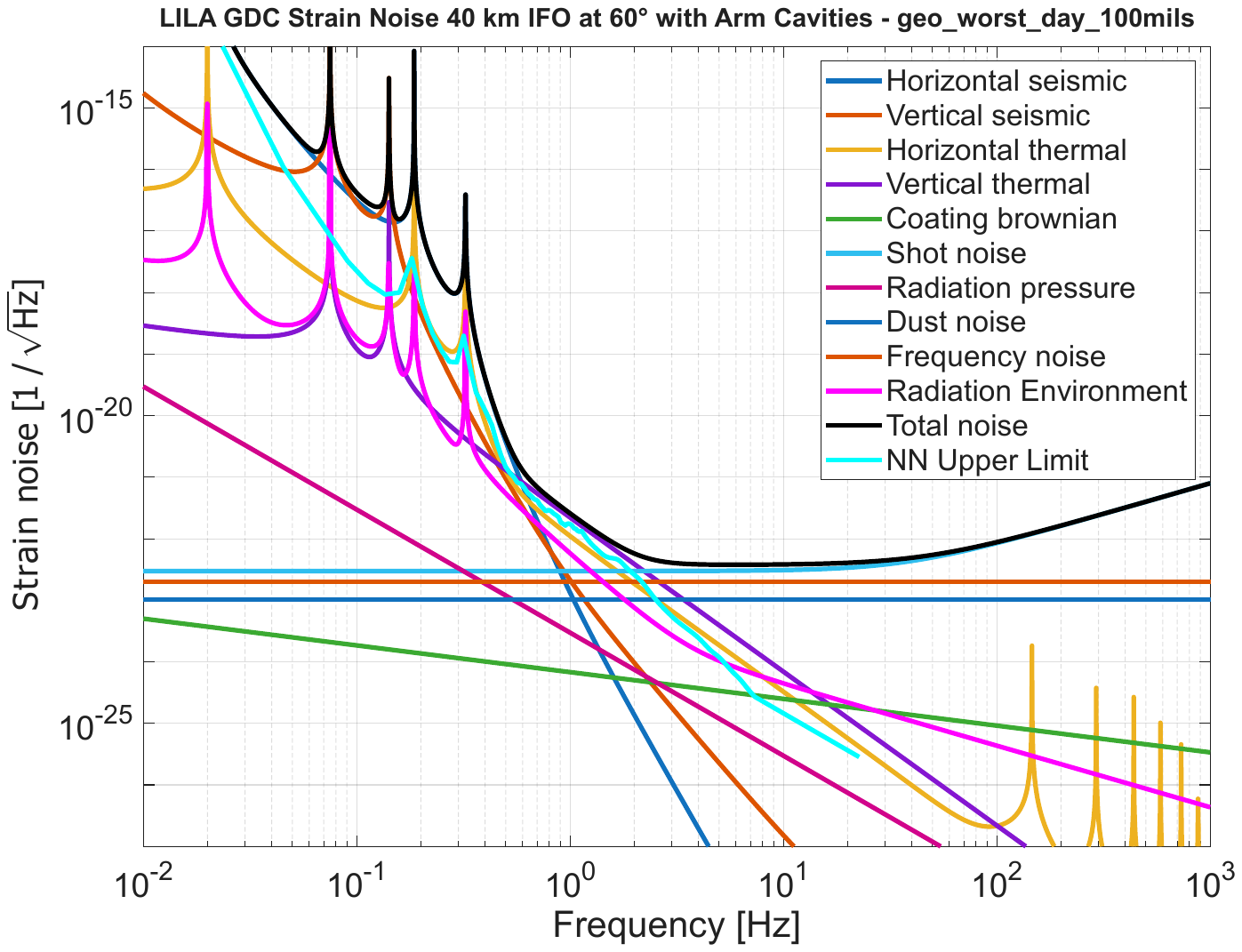}%
    \hspace*{4mm}%
}
        \caption{Worst-day}
        \label{fig:suspension-worst-day}
    \end{subfigure}

    \caption{Strain ASDs calculated using the GDC suspension transfer function from \cite{shapiro2025}. Included are the various noise sources considered in \cite{quetschke2026}. The radiation-induced curve considered here is labeled under "Radiation Environment" and is calculated assuming 100 mils shielding, electrode configuration (B), and continuous discharging. Panels (a) and (b) show the background and worst-day results, respectively. In all cases, other noise sources dominate the overall response.}
    \label{fig:suspension}
\end{figure}

 The suspension model from \cite{shapiro2025} is employed in Figure~\ref{fig:suspension} to calculate the noise on the current GDC suspension. This also includes the full LILA GDC noise spectrum from all sources considered in ~\cite{quetschke2026}. Here we choose 100 mils, electrode configuration B, and  continuous discharging as the desired GDC operating conditions. The radiation-induced noise calculated under these assumptions using the GDC suspension model is the "Radiation Environment" noise source. For both the background and worst-day environments, the radiation-induced noise is below the envelope of the other noise sources. Thus, radiation-induced noise is not expected to dominate. Final confirmation is pending a finalized configuration including the optical bench, electrode configuration, and photodiode parameters. 

\section{Conclusions}
Radiation-induced coherent and stochastic spurious signals arise from charging of the LILA Observatory GDC test mass. Continuous discharging and balanced electrode design provide options for mitigating these noise sources. Without them, coherent terms dominate and threaten the LILA measurement. With them, the overall noise is below the envelope of other sources as calculated in ~\cite{quetschke2026}. 

The final implementation of continuous discharging and the optical surfaces may introduce additional noise. To implement continuous discharging, a method such as UV irradiation must be chosen. This will induce its own technical noise, which must be considered. Dielectric patch charging and discharging must be understood as well for the finalized coating. Thermoelastic scattering coefficients and response need to be recalculated for the final TM configuration. 

These results assume the design choices outlined here. The finalized GDC design may not be able to employ these, and this may increase susceptibility to radiation-induced noise. Other combinations of mitigation strategies may need to be considered. Continuous discharge, a charge-aware electrode configuration and reasonable shielding provide complementary means to manage radiation-induced signals. The severe-event scenario may motivate temporarily restricted operation if the final sensitivity requirement is exceeded. Shielding also is highly effective at mitigating the worst-day environment. The existence of these levers and the strength of their ability to modulate the noise spectra indicate that radiation is a highly manageable source of noise. It is thus unlikely to dominate the overall response for LILA as long as charge-aware practices for enclosure and electrode design are followed.  

\appendix
\section{Additional low-energy electron emission}
\label{app:emission}
\setcounter{equation}{0}
\renewcommand{\theequation}{A.\arabic{equation}}
Low-energy electron emission accompanying a particle's approach to or departure from a material surface can contribute to test-mass charge even when those electrons are not resolved by the bulk transport calculation due to their low energies. Electronic stopping power provides a phenomenological scaling for such emission \cite{schou1980,araujo2005}. We evaluate the model at charged-particle crossings of either the optic surface or the shell's cavity-facing surface. Emission from the optic contributes positive charge while emission from the shell contributes negative. Explicit Geant4 secondary electrons remain in direct transport. The treatment does not transport these modeled low-energy electrons through the electrostatic field or determine their collection efficiency.

Electronic stopping tables are generated for the SiO$_2$--Ta$_2$O$_5$ mixture using Geant4 9.6 patch 02 with Penelope electromagnetic models. Named particles, including electrons, positrons, protons, antiprotons, muons and pions, use their own tables at their total kinetic energy at time of crossing. Other nuclei are evaluated at kinetic energy per nucleon. Values are interpolated logarithmically in both energy and stopping power, and out-of-range energies are clamped to the nearest endpoint. The grid spans $10^{-3}$--$10^5$ MeV, or MeV per nucleon for the ion tables, with 20 intervals per decade. 

At crossing $j$, the assumed mean emission multiplicity is $u_j=D_j/D_0$, where $D_j$ is the electronic stopping power in eV \AA$^{-1}$ and $D_0=10$ eV \AA$^{-1}$. This makes $u_j$ dimensionless. The normalization is an inherited phenomenological assumption from \cite{araujo2005}; it is not experimentally established for the present mixed-oxide surface. 

Let $d_h$ be an event's direct charge in elementary-charge units, $s_j=\pm1$ its emission signs, and $N_j$ independent Poisson variables conditioned on the tracked event. The event moments are
\begin{equation}
\begin{array}{ll}
N_j\sim\mathrm{Poisson}(u_j) & \langle N_j^2\rangle=u_j^2+u_j\\[2pt]
c_h=\langle n_h\mid h\rangle=d_h+\sum_j s_j u_j\\[2pt]v_h=\sum_j u_j&
\langle n_h^2\mid h\rangle=c_h^2+v_h
\end{array}
\label{eq:emission}
\end{equation}
Here $h$ labels one transported event and $j$ labels a modeled surface crossing within that event; each sum over $j$ includes all such crossings in event $h$. The random variable $N_j$ is the number of additional electrons emitted at crossing $j$, $u_j$ is its assumed Poisson mean, and $s_j=\pm1$ assigns the corresponding signed change in test-mass charge. The realized total signed event charge is $n_h=d_h+\sum_j s_jN_j$, in elementary-charge units. Thus $c_h=\langle n_h\mid h\rangle$ is the conditional mean event charge and $v_h=\langle(n_h-c_h)^2\mid h\rangle$ is its conditional variance. Angle brackets denote an expectation over the modeled Poisson emission multiplicities for a fixed transported event, and the conditioning symbol $\mid h$ indicates that the tracked event history, including $d_h$, $u_j$ and $s_j$, is held fixed.
The signs affect the conditional mean and its square, but not the positive Poisson variance. For source normalization $\mathcal K$, requested history count $N$ and importance weights $w_h$, the mean rate and physical second-moment rate are $\bar R=(\mathcal K/N)\sum_h w_hc_h$ and $\lambda_r=(\mathcal K/N)\sum_h w_h(c_h^2+v_h)$ before adding the independent allowance.

\section{Spatial thermoelastic response kernel}
\label{app:kernel}
\setcounter{figure}{0}
\renewcommand{\thefigure}{B.\arabic{figure}}
\setcounter{equation}{0}
\renewcommand{\theequation}{B.\arabic{equation}}
Yamamoto et al. derived the thermoelastic response to cosmic-ray tracks using approximately straight traversals and a representative energy loss per path length \cite{yamamoto2008}. Such approximations are useful for penetrating particles such as muon byproducts of cosmic rays considered in that work, but particles in the space environment often stop in the volume. This motivates an approach that explicitly considers the stopping. We evaluate the local deformation by relating the energy deposited to a spatial response kernel \cite{king2015thermoelastic}.

The observable is the Gaussian-weighted axial displacement of the front optical face, relative to the optic's centre of mass. An auxiliary static elastic problem applies a known virtual force $F_v=1$ N through the normalized Gaussian pressure on that face. The intensity radius is $w=0.05$ m, corresponding to a 0.10 m diameter, slightly conservative compared to Advanced LIGO's 5.3 or 6.2 cm beam spot size on the test mass mirrors ~\cite{advancedligo2015}. A uniform opposing body-force density $-F_v/V$ balances the surface load, where $V$ is the optic volume. The radial displacement is zero on the symmetry axis, all physical exterior surfaces are traction free except for the applied virtual pressure, and an arbitrary axial gauge removes rigid translation without clamping a face. 

\begin{figure}[tbp]
\centering\includegraphics[width=.92\textwidth]{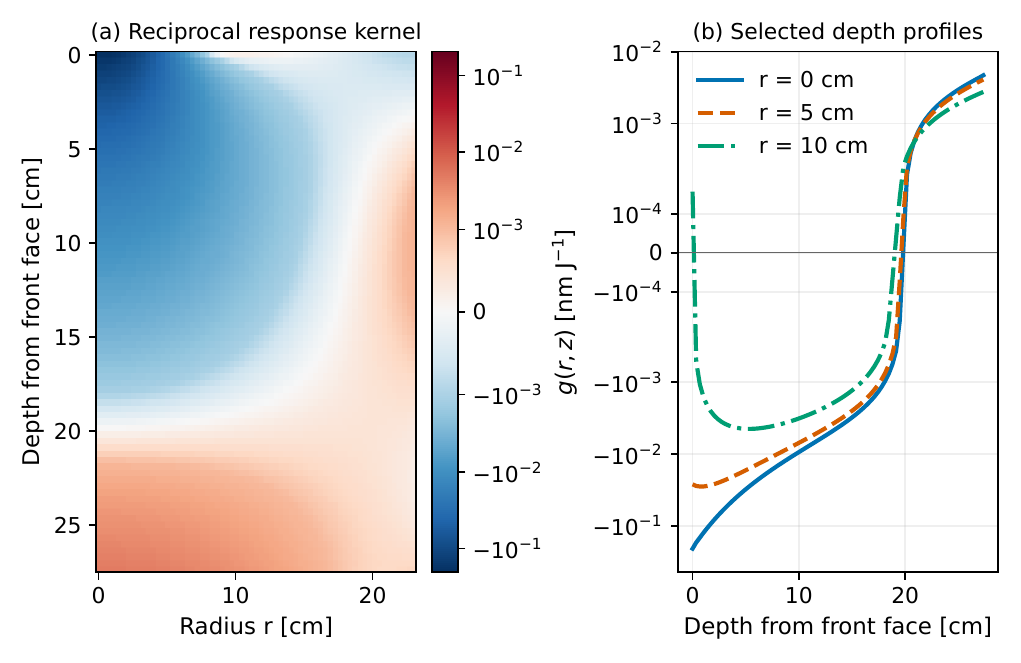}
\caption{The production thermoelastic reciprocity kernel $g(r,z)$ for the homogeneous fused-silica cylinder, shown in nm J$^{-1}$. The finite-element solution includes the centre-of-mass correction and a centred Gaussian readout with 5 cm intensity radius.}
\label{fig:kernel}
\end{figure}

Let $\mathbf u_v$ be the virtual displacement and $\theta_v=\nabla\cdot\mathbf u_v$ its volumetric strain. Elastic reciprocity then gives
\begin{eqnarray}
g(r,z)&=&\frac{Y\alpha}{(1-2\nu)\rho c_p F_v}\,\theta_v(r,z),\nonumber\\
\Delta x&=&\int_V g(r,z)\,\Delta u_{\mathrm{heat}}(r,z)\,\mathrm dV .
\label{eq:kernel}
\end{eqnarray}
Here $r$ is cylindrical radius, $z$ denotes the equivalent front-face depth, $Y$ is Young's modulus, $\nu$ is Poisson's ratio, $\alpha$ is the linear thermal expansion coefficient, $\rho$ is density, $c_p$ is specific heat and $\Delta u_{\mathrm{heat}}$ is deposited heat energy per unit volume. The kernel has units m J$^{-1}$. Its sign follows the positive axial direction toward the rear electrodes, so outward motion of the front face is negative. The resulting strain field, shown in Figure ~\ref{fig:kernel}, gives a linear mapping to energy deposition profile. During transport, charged-particle step energy is approximated as uniform along each step, integrated by two-point Gaussian quadrature on subdivisions no longer than the kernel spacing. Neutral local-interaction deposition is assigned to its interaction endpoint. 

Absolute normalization and sign are checked by uniformly heating a freely expanding cylinder. For a temperature increment $\Delta T$ and cylinder height $H$, the front face moves by $-\alpha H\Delta T/2$ relative to the centre of mass. Thus
\begin{equation}
\rho c_p\Delta T\int_V g\,\mathrm dV
=-\frac{\alpha H}{2}\Delta T .
\label{eq:uniform}
\end{equation}
For this optic the expected response is $-7.52084345\times10^{-8}$ m K$^{-1}$. Integration of the exported table agrees to relative error $1.35\times10^{-13}$; this particularly tight check verifies the global projection and normalization, not local kernel accuracy to that precision. Independent direct-heating solutions provide the more relevant spatial tests. No multiplicative adjustment was made to force agreement with the approximate Yamamoto ensemble estimate.

\section*{Data availability}
The simulation data and code supporting this study will be made
openly available through Zenodo upon publication of the article.
During peer review, materials required to evaluate and reproduce
the reported results are available privately to the editors and
reviewers on request.

\section*{Acknowledgments}
The authors acknowledge the use of OpenAI Codex for code development, proofreading, figure preparation, and reference formatting.

P.L. acknowledges the support of Labex UnivEarth (ANR-10-LABX-0023 and ANR-18-IDEX-0001) and the Universit\'e Paris Cit\'e Crossing Cutting Edges FIRE-UP project (ANR-21-EXES-0002). M.P.P. was supported by funds from the Jet Propulsion Laboratory, California Institute of Technology, under a contract with the National Aeronautics and Space Administration (80NM0018D0004). This work was supported by National Science Foundation grant NSF-2207999.  P.H., J.T., and K.J. acknowledge support from NASA grant 80NSSC26K0496 and the Vanderbilt Lunar Labs Initiative, which is supported in part by the John Templeton Foundation. J.T. and K.J. acknowledge the support from the Vanderbilt University Office of Vice-Provost for Research and Innovation. K.J. acknowledges the support of the National Science Foundation CAREER award AST-2544531.

\section*{Conflict of interest}
The authors declare no conflicts of interest.

\bibliographystyle{unsrtnat}
\bibliography{references}
\end{document}